\documentclass[aps,pra,amsmath,amssymb,amsfonts,superscriptaddress]{revtex4}
\usepackage{epsfig,graphicx,bm,pstricks,pst-node,pst-text,pst-3d,amsthm} 
\usepackage[latin1]{inputenc}

\newcommand{\beq}{\begin{equation*}}
\newcommand{\eeq}{\end{equation*}}
\newcommand{\beqa}{\begin{eqnarray*}}
\newcommand{\eeqa}{\end{eqnarray*}}

\def\deq{\begin{equation}}
\def\feq{\end{equation}}
\def\deqn{\begin{eqnarray}}
\def\feqn{\end{eqnarray}}

\def\undi{\mbox{1\hspace{-.15em}l}}

\newcommand{\ket} [1] {\vert #1 \rangle}

\newtheorem{lem}{Lemma}
\newtheorem{thm}[lem]{Theorem}
\newtheorem{defn}[lem]{Definition}

\newcommand{\abs}[1]{\left\vert#1\right\vert}

\newcommand{\protocol}[2]
{\begin{center}\framebox[1.1\width]{\begin{minipage}[l]{13cm}\begin{
center}\underline{#1}\end{center}#2
\end{minipage}}\end{center}}

\begin{document}

\title{ Simulation of bipartite qudit correlations }

\author{Julien Degorre}
\affiliation{Laboratoire de Recherche en Informatique, UMR 8263, Universit\'e Paris-Sud, 91405 Orsay, France}
\affiliation{Laboratoire d'Informatique Th\'eorique et Quantique,
D\'epartement d'Informatique et de Recherche Op\'erationnelle, Universit\'e de Montr\'eal, Canada}
\author{Sophie Laplante}
\affiliation{Laboratoire de Recherche en Informatique, UMR 8263, Universit\'e Paris-Sud, 91405 Orsay, France}
\author{J\'er\'emie Roland}
\affiliation{Quantum Information and Communication, Ecole Polytechnique, CP 165/59, 
Universit\'e Libre de Bruxelles, 1050 Bruxelles, Belgium}
\affiliation{Laboratoire de Recherche en Informatique, UMR 8263, Universit\'e Paris-Sud, 91405 Orsay, France}

\begin{abstract}

We present a protocol to simulate the quantum correlations of an
arbitrary  bipartite state, when
the parties perform a measurement according to two 
traceless binary observables. We show
that $\log(d)$ bits of classical communication is enough on average,
where $d$ is the dimension of both systems. To obtain this result, we use the
sampling approach for simulating the quantum correlations.
We discuss how to use this method in the case of qudits.

\end{abstract}

\maketitle


\section{Introduction}

In 1964 John Bell showed that the correlations exhibited by the EPR
gendanken experiment \cite{bell64,ba57} could not be  reproduced
by a so-called local hidden variable model, that is,
a model where the parties share an infinite amount of locally created
 hidden  variables. This nonlocal aspect is  one of the strangest
properties of quantum physics, and understanding this notion remains an
important problem. Recently, quantum information processing
has provided a new point of view to understand quantum nonlocality. In
particular, the framework of communication complexity has provided tools
to study nonlocality. For example, two parties, whom we call  Alice and
Bob, cannot reproduce quantum correlations if they share only hidden
random  variables (shared randomness), but in some cases, if they are
allowed to use some additional resources, it becomes possible for them
to reproduce the quantum correlations. It is precisely this amount of
additional resources which we consider here; they allow us to quantify
quantum nonlocality.

The problem of reproducing the statistics of projective measurements 
on the singlet has been widely studied, with communication as the 
additional resource. In 1992, Maudlin
\cite{maudlin92} presented a protocol in the case of mesurements 
in the real plane and proved an average-case communication upper bound 
of 1.17 bits, and independently in 1999, Brassard, Cleve and Tapp  
gave a protocol, together with a worst-case communication upper bound 
of 8 bits, for arbitrary projective measurements \cite{bct99}.
In 2000, Steiner, independently of Maudlin,
gave a protocol for projective measurements in the real plane 
with an average-case upper bound
on communication of 1.48 bits, and Cerf, Gisin and Massar~\cite{cgm00} 
proved that for an arbitrary projective measurements, $1.19$ bits of 
communication sufficed on average.
Recently, in 2003, Toner and Bacon have shown that one
bit of communication is always enough 
to reproduce the quantum
correlations for arbitrary projective measurements on the singlet 
state~\cite{tb03}.
 
Some other resources have been used to simulate quantum correlations
resulting from projective measurements on the singlet state.
These include post-selection \cite{ gg99}, and nonlocal boxes 
\cite{cgmp05}. 
In 2005, we have shown that simulating these quantum correlations
could be reduced to a sampling problem, from which we derived many of the above-mentioned protocols, in a unified framework~\cite{dlr05}.

Nevertheless, these results address the simplest scenario, that is,
simulating the correlations resulting from measurements 
on the singlet state (mostly for projective measurements, with 
a few extensions to POVMs).
There are few results about non-maximally
entangled pairs, multiparty states, higher dimensional states, or more general measurements.
One significant result in this direction is a protocol from Massar {\em et al} able to
reproduce the correlations of arbitrary measurements on any entangled pair of
$d$-dimensional states (qudits)
using $O(d \log d)$ bits of communication but no local hidden variables~\cite{mbcc00}.

In this paper, we use the sampling approach developed in \cite{dlr05},
and generalize it to the case of a bipartite pair of arbitrary-dimension
states (qudits). We study
the case where the parties make a restricted type of measurement
with only two opposite outcomes $\{1,-1\}$, that we call traceless binary
observable, or TBO.

Furthermore we impose no constraint on the bipartite
(pure) state whose correlations the parties wish to simulate; it
could be maximally entangled or non-maximally entangled.
For an arbitrary bipartite qudit pair, we show that $\log(d)$
bits of communication on average are enough
to simulate the joint correlations of the outcomes (where the joint correlation is defined as the
expectation value of the product of Alice's and Bob's outcome). In the special case of maximally
entangled qudit pairs, our protocol also reproduces the marginal probabilities, and therefore
the full probability distribution.

We will begin by describing the quantum correlations in
arbitrary dimensions that we want to simulate
classically. Then, using the sampling approach, we
will present a generalization of the local biased hidden variable model
for arbitrary dimensions, and present a classical protocol which uses
$\log(d)$ bits of communication to simulate the joint quantum correlations
of an arbitrary bipartite qudit pair.


\section{The quantum correlations}
In this section we describe the system that we want to simulate
classically using some communication.
Two parties, Alice and Bob, share an arbitrary bipartite qudit
pair. They each perform
a measurement on their part, where the measurements are
restricted to what we call traceless binary observables, described below.

We describe the measurements using observables instead of measurement
operators.
We restrict the measurements to be such that only two outcomes are 
possible,  and these outcomes
are equally  likely when the measurement is applied to 
a maximally mixed state.

\begin{defn}[Traceless Binary Observable]\label{def:TBO}
An observable
$\hat{A}$ is called a Traceless Binary Observable (TBO) if
\begin{itemize}
\item the observable is traceless (i.e.~$Tr(\hat{A}) = 0 $ )
\item the outputs of the measurement are two opposite values, more
precisely  $\hat{A}^2 = \undi $.
\end{itemize}
\end{defn}

We describe the bipartite quantum correlations on qudits
obtained when a TBO is applied to each part of the state.

\begin{defn}[Qudits TBO experiment]\label{def:bipartiteCor}
Two parties, Alice and Bob, share an arbitrary bipartite qudit pair,
$\ket{\psi}\in \mathcal{H}^d \otimes \mathcal{H}^d $. Alice and Bob
measure their part of the state according to their input, 
describing a TBO pair $\hat{A}, \hat{B}$.
They then obtain measurement outcomes $A\in\{1,-1\}$ and $B\in\{1,-1\}$
respectively.
\end{defn}

We will use the following notation throughout the paper.
Let $\mathcal{O}_d$ denote the space of TBOs over $\mathcal{H}^d$.
We use $\mathbb{S}_n$ to denote the unit hypersphere in $\mathbb{R}^{n+1}$.
(For example,  $\mathbb{S}_2$ is the unit sphere in $\mathbb{R}^{3}$.)
We will also use $\mathcal{S}_n$ to denote the surface area of $\mathbb{S}_n$.

Tsirelson showed that there is a function that maps
TBOs over the Hilbert space (matrix operators) to
points on the surface of a hypersphere, as follows~\cite{tsi85}.
This formulation of Tsirelson's theorem was pointed out to us by B.~Toner~\cite{ben}
and appears in this form in~\cite{agt06}.

\begin{thm}[Tsirelson]\label{thm:tsirelson}
For any $d>0$, and
$\ket{\psi} \in \mathcal{H}^d \otimes \mathcal{H}^d$,
there is a function $\nu:\mathcal{O}_d\longrightarrow\mathbb{S}_{2d^2-1}  $
such that the following holds.
If $\hat{A}$ and $\hat{B}$ is a TBO pair over $\mathcal{H}^d$, and $A,B$ are 
the outcomes
of measuring $\ket{\psi}$ according to $\hat{A}$ and $\hat{B}$,
then $$E(AB|\hat{A}\otimes \hat{B}) =  {\nu(\hat{A})} \cdot {\nu(\hat{B})}. $$
\end{thm}

In the remainder of the paper, we use this theorem implicitly,
and use the notation $\vec{a}=\nu(\hat{A})$,
and  $\vec{b}=\nu(\hat{B})$, to denote Alice's and Bob's inputs
(or measurement), respectively.
Furthermore, we call $E(AB|\hat{A}\otimes \hat{B})$ the joint quantum correlations.

\section{The classical protocol}
We present a protocol to simulate bipartite qudit joint quantum
correlations as defined in the previous
section, where the state $\ket{\psi}$ is an arbitrary state in 
$\mathcal{H}^d \otimes \mathcal{H}^d$. 
First, we generalize the sampling method introduced in
\cite{dlr05}.

\subsection{Local hidden biased variable model}
As in \cite{dlr05}, we consider a model where Alice and
Bob share random variables that can depend
on Alice's and/or Bob's input, which we call a local
biased random variable model.

We generalize the sampling theorem in \cite{dlr05} to arbitrary
bipartite qudit states, with TBO measurements. 
We start with a technical lemma to compute the normalization factor
of the biased distribution.

\begin{lem}
\label{lemma:Rn}
$\int_{\mathbb{S}_n} \abs{\ \vec{b}\cdot\vec{\lambda_s}}\
d\vec{\lambda_s}=  
\frac{2}{n} \mathcal{S}_{n-1}. $
\end{lem}
\begin{proof}

Since $d\vec{\lambda_s} = d\theta_{n} \ sin(\theta_{n-1})
d\theta_{n-1} \ sin^{2}(\theta_{n-2}) d\theta_{n-2} ...
sin^{n-1}(\theta_{2}) d\theta_{2} \  sin^{n-1}(\theta_{1}) d\theta_{1}
$, 
\beqa
\lefteqn{\int_{\mathbb{S}_n} \abs{\ \vec{b}\cdot\vec{\lambda_s}}\
d\vec{\lambda_s}}\\
 &=& \int_{0}^{2\pi} d\theta_{n}
\int_{0}^{\pi} sin(\theta_{n-1}) d\theta_{n-1} \int_{0}^{\pi}
sin^{2}(\theta_{n-2}) d\theta_{n-2}...
\int_{0}^{\pi} sin^{n-2}(\theta_{2}) d\theta_{2}
\int_{0}^{\pi} sin^{n-1}(\theta_{1}) \abs{cos(\theta_{1})}\ d\theta_{1}
\\
&=& \mathcal{S}_{n-1}\ 2\ \int_{0}^{\pi/2} sin^{n-1}(\theta_{1})
cos(\theta_{1})\ d\theta_{1}
=2\ \mathcal{S}_{n-1} [sin^n(\theta_1)/n]^{\pi/2}_{0}\\
&=&\frac{2}{n} \ \ \mathcal{S}_{n-1}. 
\eeqa

\end{proof}

We will write $R_n = \frac{2}{n}  \mathcal{S}_{n-1}$ to simplify notation.

\begin{thm}[Generalized sampling theorem]\label{thm:Genesampling}
Let $\vec{a}$ and $\vec{b}$ $\in \mathbb{S}_n$ be Alice's and Bob's
inputs.
If Alice and Bob share a random variable
$\vec{\lambda_s}\in\mathbb{S}_n$ distributed according to a biased
distribution with probability density
\beq
\rho(\vec{\lambda_s}|\vec{a}\vec{b})=\rho_{\vec{a}}(\vec{\lambda_s})
= \frac{ \abs{\ \vec{a}\cdot\vec{\lambda_s}}}{R_n},
\eeq
then they can simulate the joint correlations
$$E(AB|\vec{a}\vec{b}) =  \vec{a}\cdot\vec{b},$$
with marginal expectations
$$ E(A|\vec{a}\vec{b})=E(B|\vec{a}\vec{b})=0.$$
\end{thm}

This says that in this model, Alice and Bob can simulate these correlations
without any further resource, that is, simulating the bipartite two
output joint quantum correlations reduces to distributed sampling from
the distribution $\rho_{\vec{a}}$.

\begin{proof}
Consider the protocol where  Alice and Bob set their respective outputs to
$A(\vec{a}\vec{\lambda_s})=\textrm{sgn}(\vec{a}\cdot\vec{\lambda_s})$
and
$B(\vec{b}\vec{\lambda_s})=\textrm{sgn}(\vec{b}\cdot\vec{\lambda_s})$,
where $\textrm{sgn}(x)=1$ for $x\geq 0$
and
$\textrm{sgn}(x)=-1$ for $x<0$ $(x\in\mathbb{R})$.
Then the joint expectation $E(AB|\vec{a}\vec{b})$ is given by

\beqa
E(AB|\vec{a}\vec{b})&=&
\int_{\mathbb{S}_n} \rho_{\vec{a}}(\vec{\lambda_s})\
A(\vec{a},\vec{\lambda_s})B(\vec{b},\vec{\lambda_s})\ d\vec{\lambda_s}\\ 
&=&
\frac{1}{R_n}
\int_{\mathbb{S}_n}
\abs{\vec{a}\cdot\vec{\lambda_s}}\
\textrm{sgn}(\vec{a}\cdot\vec{\lambda_s})\
\textrm{sgn}(\vec{b}\cdot\vec{\lambda_s})\ d\vec{\lambda_s}\\
&=&
\frac{1}{R_n}\int_{\mathbb{S}_n}
(\vec{a}\cdot\vec{\lambda_s})\
\textrm{sgn}(\vec{b}\cdot\vec{\lambda_s})\ d\vec{\lambda_s}\\
&=&\frac{1}{R_n}
\vec{a}\cdot ( \int_{\mathbb{S}_n}
\vec{\lambda_s}\ \textrm{sgn}(\vec{b}\cdot\vec{\lambda_s})\
d\vec{\lambda_s}).\label{frac}
\eeqa

Observe that the final integral is invariant by rotation around $\vec{b}$,
so it must be the case that 
\begin{equation}
\label{eq:constante}
\int_{\mathbb{S}_n}  \vec{\lambda_s}\ \textrm{sgn}(\vec{b}\cdot\vec{\lambda_s})\ d\vec{\lambda_s} = c\; \vec{b},
\end{equation}

with $c$ a real constant.

Multiplying Equation~\ref{eq:constante} by $\vec{b}$ on either side to compute 
the constant, we obtain
$$\int_{\mathbb{S}_n}
\vec{b} \cdot \vec{\lambda_s}\
\textrm{sgn}(\vec{b}\cdot\vec{\lambda_s})\
d\vec{\lambda_s}
= c ~(\vec{b} \cdot \vec{b}) = c. $$

By Lemma~\ref{lemma:Rn}, $c=R_n$, therefore, 
$$ E(AB|\vec{a}\vec{b})=  \frac{R_n}{R_n} ~\vec{a}\cdot \vec{b} =  \vec{a}\cdot \vec{b} .
$$

Finally, we compute the marginal distributions.
It is easy to see that 
\begin{eqnarray*}
 E(A|\vec{a}\vec{b})&=&
\int_{\mathbb{S}_n} \rho_{\vec{a}}(\vec{\lambda_s})\ A(\vec{a},\vec{\lambda_s})\ d\vec{\lambda_s}\\
&=& 
\frac{1}{R_n} \int_{\mathbb{S}_n} \abs{\vec{a}\cdot\vec{\lambda_s}}\ \textrm{sgn}(\vec{a}\cdot\vec{\lambda_s})\ d\vec{\lambda_s}\\
&=&
\frac{1}{R_n} \int_{\mathbb{S}_n} (\vec{a}\cdot\vec{\lambda_s})\ d\vec{\lambda_s} = 0.
\end{eqnarray*} 

For the second marginal, 
$$
 E(B|\vec{a}\vec{b}) = \int_{\mathbb{S}_n} \rho_{\vec{a}}(\vec{\lambda_s})\ B(\vec{b},\vec{\lambda_s})\ d\vec{\lambda_s} 
= \frac{1}{R_n} \int_{\mathbb{S}_n} \abs{\vec{a}\cdot\vec{\lambda_s}}\ \textrm{sgn}(\vec{b}\cdot\vec{\lambda_s})\ d\vec{\lambda_s}.
$$
With $\mathbb{S}_{+}$, $\mathbb{S}_{-}$ the 
half-spheres with respect to $\vec{a}$,
and with $\vec{\lambda}_+ \in \mathbb{S}_{+}$ and 
$\vec{\lambda}_- \in \mathbb{S}_{-}$,
\begin{equation}
\nonumber
E(B|\vec{a}\vec{b}) = 
\frac{1}{R_n} \int_{\mathbb{S}_{+}} \vec{a}\cdot\vec{\lambda_+}\ \textrm{sgn}(\vec{b}\cdot\vec{\lambda_+})\ d\vec{\lambda_+}
-\frac{1}{R_n} \int_{\mathbb{S}_{-}} \vec{a}\cdot\vec{\lambda_-}\ \textrm{sgn}(\vec{b}\cdot\vec{\lambda_-})\ d\vec{\lambda_-}.
\end{equation}
We make a variable substitution 
$\vec{\lambda}_- = - \vec{\lambda}_+ $ and we obtain
\begin{equation}
\nonumber
E(B|\vec{a}\vec{b}) = 
\frac{1}{R_n} \int_{\mathbb{S}_{+}} \vec{a}\cdot\vec{\lambda_+}\ \textrm{sgn}(\vec{b}\cdot\vec{\lambda_+})\ d\vec{\lambda_+}
-\frac{1}{R_n} \int_{\mathbb{S}_{+}} \vec{a}\cdot\vec{\lambda_+}\ \textrm{sgn}(\vec{b}\cdot\vec{\lambda_+})\ d\vec{\lambda_+} = 0.
\end{equation}
\end{proof}

Note that we obtain $E(A|\vec{a}\vec{b})=E(B|\vec{a}\vec{b})=0$. 
In the case of measurements on maximally entangled states, it
is also the case that the marginal expectations are zero. 
This is because the reduced states of Alice and Bob are maximally 
mixed and, for traceless binary observables, the marginal 
distributions are uniform, 
so that $E(A|\vec{a}\vec{b})=E(B|\vec{a}\vec{b})=0$.
However, for arbitrary states,
our method will reproduce the joint quantum correlations 
$E(AB|\vec{a}\vec{b})$,
but not necessarily the marginal distributions. 

\subsection{Sampling the biased distribution: the rejection method}
It now remains to show how Alice and Bob can obtain a shared
sample $\vec{\lambda_s} \in  \mathbb{S}_n$  distributed according to
 the biased distribution
$\rho_{\vec{a}}(\vec{\lambda_s})
=  \abs{\ \vec{a}\cdot\vec{\lambda_s}} /{R}_{n}$, with help of shared $\vec{\lambda}$ uniformly distributed on $\mathbb{S}_n$, and with the additional
help of communication.
Using the same idea as \cite{ steiner00,feldmann95,dlr05}, 
Alice uses the rejection method to perform the
sampling.

\begin{thm}
There is a local hidden variable protocol that simulates
the joint correlations of TBO measurements on a bipartite pair of $d$-dimensional 
states
using $\log(d)+O(1)$ bits of communication on average.
\end{thm}

As noted above, for arbitrary states, we reproduce the joint correlations
$E(AB|\vec{a}\vec{b})$, and in the special case of maximally entangled 
states, we reproduce the full joint distribution exactly.

\begin{proof}
By Theorem~\ref{thm:Genesampling}, it suffices to give a protocol
to sample the distribution $\rho_{\vec{a}}$ on $\mathbb{S}_n$ for $n=2d^2-1$ in a distributed fashion.
We show that this can be achieved with $\log(d)$ communication on average.
We obtain a sample by applying the rejection method \cite{devroye},
using unbiased (uniform) shared random variables.

Let $U(\vec{\lambda})$ be the uniform probability density function on
$\mathbb{S}_n$, that is, $U(\vec{\lambda})= 1/\mathcal{S}_n$.
The protocol is as follows. 
\begin{enumerate}
\item Alice obtains a uniform sample $\vec{\lambda} \sim U$
\item She computes $|\vec{a}\cdot\vec{\lambda}|$, and accepts $\vec{\lambda}$
with the corresponding probability.  If she accepts, she sends
Bob the iteration at which this occurred.
\item If she rejects, she starts over with a new sample $\vec{\lambda}$.
\end{enumerate}

We compute the probability that Alice accepts at a given iteration
(let us call this event ``ok''), on average over the choice of $\vec{\lambda}$.
\beqa
p(\textrm{ok})=\int_{\mathbb{S}_n} p(\textrm{ok}|\vec{\lambda}) \rho(\vec{\lambda})\ d\vec{\lambda}.
\eeqa
Since $\vec{\lambda}$ is uniformly distributed on $\mathbb{S}_n$, we have
$\rho(\vec{\lambda})=U(\vec{\lambda})= 1/\mathcal{S}_n$. Moreover, we accept a given
$\vec{\lambda}$ with probability $p(\textrm{ok}|\vec{\lambda})=|\vec{a}\cdot\vec{\lambda}|$, so that
\begin{eqnarray}
p(\textrm{ok})&=&\frac{1}{\mathcal{S}_n}\int_{\mathbb{S}_n}|\vec{a}\cdot\vec{\lambda}|\ d\vec{\lambda}
= \frac{\mathrm{R}_n}{\mathcal{S}_n},\label{probRej}
\end{eqnarray}
where we have used Lemma~\ref{lemma:Rn}.

We may now compute the distribution of accepted $\vec{\lambda}$'s as follows
\beqa
\rho(\vec{\lambda}|\textrm{ok})&=&\frac{\rho(\vec{\lambda})\ p(\textrm{ok}|\vec{\lambda})}{p(\textrm{ok})}\\
&=&\frac{|\vec{a}\cdot\vec{\lambda}|}{\mathrm{R}_n},
\eeqa
which corresponds to $\rho_{\vec{a}}$ as required. This proves that the protocol achieves its goal.
It remains to show that it requires at most $O(\log(d))$ bits of communication on average.

The message sent in the protocol is the
iteration $i$ at which Alice accepts the current
uniform sample.   The distribution of the messages
behaves according to a Poisson distribution $P_p$, that is,
$i$ is sent with probability
\beq
P_p(i) = (1-p)^{i{-}1}p,
\eeq
where $p=p(\textrm{ok})$ in our case.
To compute the number of bits sent on average, it suffices
to compute the entropy of this distribution, which we do below.
From Equation~\ref{probRej} and Lemma~\ref{lemma:Rn}, we have
\beqa
p(\textrm{ok})
&=&\frac{2 \ \mathcal{S}_{n-1}}{n\ \mathcal{S}_n}.
\eeqa
The surface area $\mathcal{S}_n$ of the hyper-sphere $\mathbb{S}_n$ is given by
\beq
\mathcal{S}_n= \frac{2 \pi^{\frac{n+1}{2}}}{\Gamma(\frac{n+1}{2})}, 
\eeq
where $\Gamma$ is the well known gamma function, defined as
\beq
\Gamma(x)=\int_0^\infty t^{x-1} e^{-t} dt.
\eeq
The acceptance probability is
\beq
p(\textrm{ok})=\frac{2}{n \sqrt{\pi}} \frac{ \Gamma(\frac{n+1}{2})}{\Gamma(\frac{n}{2})}.
\eeq
Using the fact that \cite[Ex.~9.44]{gkp94}
$$\frac{ \Gamma(\frac{n+1}{2})}{\Gamma(\frac{n}{2})} = \sqrt{\frac{n}{2}}+ O(\frac{1}{\sqrt{n}})$$
and in particular for any $n\geq 1$,
\beq
\frac{1}{2}\sqrt{\frac{n}{2}}\leq \frac{ \Gamma(\frac{n+1}{2})} {\Gamma(\frac{n}{2})}  \leq\sqrt{\frac{n}{2}}.
\eeq
Therefore
\beq
\sqrt{\frac{1}{2 \pi n}}\leq p(\textrm{ok})\leq   \sqrt{\frac{2}{\pi n}}.
\eeq
The entropy of the messages is given by 
\begin{eqnarray*}
H(P_p) &=&\sum_i P_p(i)\ \log\left(\frac{1}{P_p(i)}\right)\\
&=&\log\left(\frac{1}{p}\right) + \frac{1-p}{p} \log \left(\frac{1}{1-p}\right).
\end{eqnarray*}
So in our case, we get
\beqa
 H(P_p)
&\leq& \log(\sqrt{2 \pi n})
 + \frac{1-\sqrt{\frac{1}{2 \pi n}}}{\sqrt{\frac{1}{2 \pi n}}} \log\left(\frac{1}{1-\sqrt{\frac{2}{\pi n}}}\right)\\
&\leq&\frac{1}{2}\log(n) + O(1).
\eeqa
So, with $n=2 d^2-1$, $\log(d)+O(1)$ bits of communication are sufficient on 
average to simulate the joint quantum correlations of an arbitrary bipartite qudit pair, measured according to a TBO. 
\end{proof}

\section{Discussion and conclusion}

We have shown that in the general case of bipartite qudit pairs, 
we can apply the sampling approach of \cite{dlr05} to simulate 
the joint quantum correlations, using $\log(d)$ 
bits of communication on average.

There are very few results in the literature concerning 
settings that are more general than projective measurements 
on maximally entangled qubit pairs.
In the case of qudits, Bacon and Toner have shown how to simulate 
joint quantum correlations arising from TBO measurements with constant communication on average, but the 
correlations are simulated approximately \cite{ben}, whereas here we simulate the correlations exactly. 
In \cite{mbcc00}, Massar {\em et al} gave a protocol that simulates the correlations of any local measurement
on an arbitrary bipartite state exactly, but within a different model, which uses communication only and
no local hidden variables.

In \cite{dlr05}, we considered two ressources other than
communication: post-selection, and nonlocal boxes.
In the qubit setting, instead of iterating the rejection method
until a suitable sample was selected, it could be proven that
if the first sample failed the selection, then the second sample 
could be used.  In this case, one bit of communication sufficed 
for Alice to communicate to Bob which sample to use, and the 
method could also be adapted to obtain a protocol that makes a single use of
a nonlocal box.

It turns out that this so-called ``choice method'' does not
extend directly to dimensions other than two.  Therefore, we do not
obtain a worst case analysis in this more general setting, 
nor do we get a protocol using nonlocal boxes.

On the other hand, our analysis immediately applies to 
protocols using post-selection, that is, where the protocol
is allowed to abort with some probability.  Here, the
protocol succeeds with probability at least $\sqrt{\frac{1}{2\pi n}}$,
where $n=2d^2-1$, that is, $O(\frac{1}{d})$.

\begin{acknowledgements}
We would like to thank Ben Toner for many fruitful discussions.
This work has been supported by the European Commission under the
Integrated Project Qubit Applications (QAP) funded by the IST
directorate as Contract Number 015848, and the ANR Blanc AlgoQP.
J. R. acknowledges support from the Belgian FNRS.
\end{acknowledgements}

\bibliography{maBiblio}

\end{document}